\documentclass[oneside,reqno]{amsart}
\usepackage{amsmath}
\usepackage{amssymb}
\usepackage{xspace}
\usepackage[mathscr]{eucal}

\newtheorem{theorem}{Theorem}
\newtheorem{lemma}[theorem]{Lemma}
\theoremstyle{remark}
\newtheorem*{remark}{Remark}
\newcommand{\mat}[3]{\ensuremath{
																								\left \langle  \vphantom{#2 #3}   #1   
                        \right|    					\, #2\,   
                        \left|    \vphantom{#2 #1} #3   
                        \right \rangle
                        								}
                     }
\newcommand{\bmat}[3]{\ensuremath{
																								\bigl \langle     #1   \bigr|    \, #2\,   \bigl|     #3   \bigr \rangle
                        									}
                     }
\newcommand{\submatB}[5]{\ensuremath{     {\vphantom{\ket{#1}}}_{#4} \mspace{- #5 mu} 
																																														 \mat{#1}{#2}{#3}
																																								}
                           } 

\newcommand{\ket}[1]{\ensuremath{		\left| #1 \right> 
																																			  }
																									}
\newcommand{\bket}[1]{\ensuremath{		\bigl| #1 \bigr> 
																																			  }
																									}
\newcommand{\bra}[1]{\ensuremath{		\left< #1 \right| 
																																			  }
																									} 
\newcommand{\bbra}[1]{\ensuremath{		\bigl< #1 \bigr| 
																															  }
																					}
\newcommand{\coh}[3]{\ensuremath{		\left( #1, #2 \right)_{#3} 
																																			  }
																									}

\newcommand{\overlap}[2]{\ensuremath{ 
																								\left \langle    #1 \vphantom{#2 } \,
                        \right| \left.   #2 \vphantom{#1}
                        \right \rangle
                        									}
                     }
\newcommand{\boverlap}[2]{\ensuremath{ 
																								\bigl \langle #1 \, \bigr| \bigl. #2 \bigr \rangle
                        									}
                     }
\newcommand{\suboverlapB}[4]{\ensuremath{     {\vphantom{\ket{#1}}}_{#3} \mspace{- #4 mu} 
																																														 \overlap{#1}{#2} 
																																								}
                           }      

\newcommand{\comm}[2]{\ensuremath{  \left[ #1, #2 \right] }}

\newcommand{\pt}{   \ensuremath{     \phi_{\mathrm{ap}}     }}
\newcommand{\xOp}{ \ensuremath{  \hat{x}  }}
\newcommand{\pOp}{ \ensuremath{  \hat{p}  }}
\newcommand{\xiOp}{ \ensuremath{  \hat{x}_{\mathrm{i}}  }\xspace}
\newcommand{\piOp}{ \ensuremath{  \hat{p}_{\mathrm{i}}  }\xspace}
\newcommand{\xfOp}{ \ensuremath{  \hat{x}_{\mathrm{f}}  }\xspace}
\newcommand{\pfOp}{ \ensuremath{  \hat{p}_{\mathrm{f}}  }\xspace}
\newcommand{\xf}{   \ensuremath{   x_{\mathrm{f}}  } }

\newcommand{\MxOp}{  \ensuremath{   \hat{\mu}_{\mathrm{X}}    } }

\newcommand{\MxfOp}{  \ensuremath{   \hat{\mu}_{\mathrm{Xf}}    } }
\newcommand{\MpfOp}{  \ensuremath{   \hat{\mu}_{\mathrm{Pf}}    } }
\newcommand{\Mx}{   \ensuremath{   \mu_{\mathrm{X}}  } }

\newcommand{\Mxf}{   \ensuremath{   \mu_{\mathrm{Xf}}  } }
\newcommand{\Mpf}{   \ensuremath{   \mu_{\mathrm{Pf}}  } }

\newcommand{\yfOp}[1]{  \ensuremath{ \hat{y}_{\mathrm{f} #1 } } }

\newcommand{\yfv}[1]{  \ensuremath{ y_{\mathrm{f} #1 } } }

\newcommand{\Exi}{\ensuremath{       \hat{\epsilon}_{\mathrm{Xi}}       }\xspace}
\newcommand{\Epi}{\ensuremath{       \hat{\epsilon}_{\mathrm{Pi}}       }\xspace}

\newcommand{\Exf}{\ensuremath{       \hat{\epsilon}_{\mathrm{Xf}}       }\xspace}
\newcommand{\Epf}{\ensuremath{       \hat{\epsilon}_{\mathrm{Pf}}       }\xspace}

\newcommand{\Exfv}{\ensuremath{       \epsilon}_{\mathrm{Xf}       }\xspace}

\newcommand{\RErr}{\ensuremath{     \Delta_{\mathrm{ei}}  }\xspace}
\newcommand{\PErr}{\ensuremath{     \Delta_{\mathrm{ef}}  }\xspace}

\begin{document}
\begin{titlepage}
\begin{center}
\bfseries
OPTIMAL JOINT MEASUREMENTS OF POSITION AND MOMENTUM
\end{center}
\vspace{1 cm}
\begin{center}
D M APPLEBY
\end{center}
\begin{center}
Department of Physics, Queen Mary and
		Westfield College,  Mile End Rd, London E1 4NS, UK
\end{center}
\vspace{0.3 cm}
\begin{center}
(e-mail:  D.M.Appleby@qmw.ac.uk) 
\end{center}
\vspace{1.5 cm}
\begin{center}
\textbf{Abstract}\\
\vspace{0.35 cm}
\parbox{10.5 cm }{   The distribution of  measured values for maximally accurate, unbiased simultaneous 
                     measurements of position and momentum is investigated.  It is shown, that if the measurement 
                     is retrodictively optimal, then the distribution of results is given by the initial state 
                     Husimi function (or $Q$-representation).  If the measurement is predictively optimal, 
                     then the distribution of results is related to the final state anti-Husimi function 
                     (or $P$-representation).  The significance of this universal property for the interpretation 
                     of the Husimi function is discussed. 
                 }\\     
\vspace{1.0 cm}
\parbox{10.5 cm }{
}
\end{center}
\vspace{2 cm}
\begin{center}
Report no. QMW-PH-98-14
\end{center}
\end{titlepage}
\section{Introduction}
There is currently some interest in simultaneous measurements 
of position and momentum~\cite{Arthurs,Halliwell,LeonOth,Torma,Power,Wodk,Ban,LeonBook,BuschBook}.  
Measurements of this kind have an immediate, technical relevance to the 
field of quantum optics.  They also have a rather more general, conceptual relevance to the 
problem of understanding the classical limit.

In two previous papers~\cite{self1,self2} we discussed the accuracy
of such measurements.  We began with  Braginsky and Khalili's analysis~\cite{Braginsky} of single
measurements of $x$ only, and extended it to the case of simultaneous measurements
of $x$ and $p$ together.  We identified two types of error:  the retrodictive (or determinative)
errors  
$\RErr x$, $\RErr p$; and the predictive (or preparative) errors $\PErr x$, $\PErr p$.  
We showed, that
subject to some rather unrestrictive assumptions regarding the nature of the measurement
process, they satisfy the retrodictive error relationship
\begin{equation*}
   \RErr x \, \RErr p  \ge \frac{\hbar}{2}
\end{equation*}
and the predictive error relationship
\begin{equation*}
   \PErr x \, \PErr p  \ge \frac{\hbar}{2}
\end{equation*}
In the following we address the question:  what (if anything) can be said about the distribution of 
measured values in those cases where the lower bound set by one of these inequalities
is actually achieved?  

We begin, in Section~\ref{sec:  RetroOpt}, by considering measurements which are retrodictively 
optimal.  We define a retrodictively optimal measurement to be any measurement
belonging to the class of processes defined in ref.~\cite{self2} 
which minimises the product of retrodictive errors 
(so that $\RErr x \, \RErr p = \frac{\hbar}{2}$), and
which is retrodictively unbiased [so that the systematic errors of retrodiction
are zero---see Eq.~(\ref{eq:  RetOptCond1}) below].  We show, that for such 
measurements, the distribution of measured values is always given by the initial system state
Husimi function~\cite{Husimi,Reviews}.
This result is the extension, to the general class of measurement 
processes defined in ref.~\cite{self2}, of the result proved 
by Ali and Prugove\v{c}ki~\cite{BuschBook,Prugovecki} for the case of 
measurement processes which are Galilean covariant, and (using rather different methods) in 
ref.~\cite{self3} for the particular case of the Arthurs-Kelly process.

A number of related results have been obtained by other authors.
In the case of the Arthurs-Kelly process
several authors~\cite{Arthurs,LeonBook} have shown, that the Husimi function describes the
distribution of measured values for certain choices of initial apparatus state.
Leonhardt and Paul~\cite{LeonOth} have shown that the same is true for a number of other 
processes.  However, these authors all confine themselves to
particular examples of simultaneous measurement processes.  They do not consider
measurement processes in general.  Moreover, they do not relate the distribution of 
measured values to the accuracy 
of the measurement process.
In particular, they do not show that the Husimi function describes the distribution
of results in precisely those cases where the measurement is retrodictively
``optimal'' or ``best''.

W\'{o}dkiewicz has proposed an operational approach to the problem of 
phase space measurement~\cite{Wodk,Ban}.  
If one takes the filter reference state (or ``quantum ruler'') 
used to define his operational distribution to be a squeezed vacuum state,
and a minimum uncertainty state for $\xOp$ and $\pOp$, then one obtains the
Husimi function.  It could be said that the Husimi function is the operational
distribution corresponding to the case when the quantum ruler is most exactly and finely
calibrated---a fact which obviously ties in with the result which we
prove in Section~\ref{sec:  RetroOpt}
below.

However, the result which is most similar to ours is the one
obtained by Ali and 
Prugove\v{c}ki~\cite{BuschBook,Prugovecki}, working within the framework
of the approach based on POVM's (positive operator valued measures) and unsharp 
observables.  In fact, their result is the same as ours, except that we
prove it under much less restrictive conditions (unlike Ali and Prugove\v{c}ki
we do not assume Galilean covariance. Galilean 
covariance is a consequence of the result which we prove, not a presupposition).
It may also be worth remarking that our way of analysing the concept
of a simultaneous measurement process is rather different from 
theirs.  In particular, the objections recently raised by
Uffink~\cite{Uffink} do not apply to our arguments.

In Section~\ref{sec:  PreOpt} we go on to consider predictively optimal measurements---\emph{i.e.}
measurements of the type defined in ref.~\cite{self2} which minimise
the product of predictive errors (so that $\PErr x \, \PErr p = \frac{\hbar}{2}$).
We show, that in the case of such a measurement, the distribution of results is related
to the final state anti-Husimi function~\cite{Reviews,PRep} (the $P$-function of quantum optics).
This result also represents an extension, to the general class of measurement
processes defined in ref.~\cite{self2}, of a result proved in 
ref.~\cite{self3}, for the special case of the Arthurs-Kelly process.

In Section~\ref{sec:  Interp} we conclude by discussing the bearing of our results
on the
interpretation of the Husimi function.  In 
Section~\ref{sec:  RetroOpt} we show that 
the Husimi function describes the outcome of \emph{any} retrodictively optimal process.  
In other words, the Husimi
function has a universal significance.  We will argue that this lends some support to the 
idea, that the Husimi function is the quantum mechanical entity which 
most nearly resembles the classical concept, of the ``real'' or ``objective''
distribution describing an ensemble of identically prepared systems.
\section{Retrodictively Optimal Measurements}
\label{sec:  RetroOpt}
We will say that a simultaneous measurement process of the kind defined in 
ref.~\cite{self2} is \emph{retrodictively  optimal} if
\begin{enumerate}
\item  The process is retrodictively unbiased, so that
       \begin{equation}
           \mat{\psi \otimes \pt}{\Exi}{\psi \otimes \pt}
       =   \mat{\psi \otimes \pt}{\Epi}{\psi \otimes \pt}
       =   0
       \label{eq:  RetOptCond1}
       \end{equation}
       for all $\ket{\psi}\in \mathscr{H}_{\mathrm{sy}}$.
\item  The product of retrodictive errors achieves its lower bound, so that
       \begin{equation}
           \RErr x \, \RErr p  = \frac{\hbar}{2}
       \label{eq:  RetOptCond2}
       \end{equation}
\end{enumerate}
Here and in the sequel we employ the notation and terminology of ref.~\cite{self2}.
Thus, $\ket{\psi} \in \mathscr{H}_{\mathrm{sy}}$ and
$\ket{\pt} \in \mathscr{H}_{\mathrm{ap}}$ are the initial states of the system and apparatus
respectively.  
$\Exi$, $\Epi$ are the retrodictive error operators.  $\RErr x$, $\RErr p$
are the maximal rms errors of retrodiction.

In ref~\cite{self3} we considered the special case of 
the Arthurs-Kelly process.  In that case one has the commutation relation
\begin{equation}
    \comm{\Exi}{\Epi} = - i \hbar
\label{eq:  RetErrComRel}
\end{equation}
This relationship, and the condition of 
Eq.~(\ref{eq:  RetOptCond2}), together imply
Eq.~(\ref{eq:  RetOptCond1}).  In the general case, however, it is necessary
to impose the requirement, that the measurement be retrodictively unbiased, as a separate
condition.

In the general case the case the commutation relationship of
Eq.~(\ref{eq:  RetErrComRel}) cannot be assumed. However, it was shown
in ref.~\cite{self2} that Eq.~(\ref{eq:  RetOptCond1}) implies
the  weaker statement
\begin{equation}
    \mat{\psi \otimes \pt}{\comm{\Exi}{\Epi}}{\psi \otimes \pt} = - i \hbar
\label{eq:  RetOptComExpect}
\end{equation}
for every normalised $\ket{\psi}\in \mathscr{H}_{\mathrm{sy}}$ [but fixed $\ket{\pt}$]. 
It turns out that this is enough to prove, that the distribution
of measured values is given by the initial system state Husimi function,
for any retrodictively optimal process.  However, the fact that we can no longer assume
the commutation relationship of Eq.~(\ref{eq:  RetErrComRel}),
means that the proof of this statement is less straightforward
than the proof given in ref.~\cite{self3}, for the special case of the
Arthurs-Kelly process.

In view of 
Eqs.~(\ref{eq:  RetOptCond2}) and~(\ref{eq:  RetOptComExpect}) we have
\begin{equation}
    \mat{\psi \otimes \pt}{\Exi^2}{\psi \otimes \pt} \,
    \mat{\psi \otimes \pt}{\Epi^2}{\psi \otimes \pt}
=   \frac{\hbar^2}{4}
\label{eq:  RetOptUniBnd}
\end{equation}
for every normalised $\ket{\psi}\in \mathscr{H}_{\mathrm{sy}}$.  We 
deduce:
\begin{lemma}
\label{lem:  RetRes}
Given any retrodictively optimal measurement process
with initial apparatus state $\ket{\pt}$, there exists a fixed number
$\lambda_{\mathrm{i}}$ such that
\begin{equation*}
\begin{split}
    \mat{\psi \otimes \pt}{\Exi^2}{\psi \otimes \pt} & = \frac{\lambda_{\mathrm{i}}^2}{2} \\
    \mat{\psi \otimes \pt}{\Epi^2}{\psi \otimes \pt} & = \frac{\hbar^2}{2 \lambda_{\mathrm{i}}^2}
\end{split}
\end{equation*}
for every normalised $\ket{\psi}\in \mathscr{H}_{\mathrm{sy}}$.  
\end{lemma}
\begin{remark}
We will refer to $\lambda_{\mathrm{i}}$ as the retrodictive spatial resolution of 
the measurement.
\end{remark}
\begin{proof}
For each normalised $\ket{\psi}\in \mathscr{H}_{\mathrm{sy}}$ define the number $\lambda_{\psi}$ by
\begin{equation*}
     \lambda_{\psi}
=    \left( 2 \mat{\psi \otimes \pt}{\Exi^2}{\psi \otimes \pt} \right)^{\frac{1}{2}}
\end{equation*}
In view of Eq.~(\ref{eq:  RetOptUniBnd}) we then have
\begin{equation*}
     \left( \mat{\psi \otimes \pt}{\Epi^2}{\psi \otimes \pt} \right)^{\frac{1}{2}}
=    \frac{\hbar}{\sqrt{2} \lambda_{\psi}}
\end{equation*}
We have from the definitions~\cite{self2} of $\RErr x$, $\RErr p$
\begin{align*}
    \RErr x  
& = \sup_{\ket{\psi} \in \mathscr{S}}
       \Bigl(\mat{\psi \otimes \pt}{\Exi^2}{\psi \otimes \pt} \Bigr)^{\frac{1}{2}} 
  = \frac{ \sup_{\ket{\psi} \in \mathscr{S}} \left(\lambda_{\psi}\right) }{\sqrt{2}}
\\
\intertext{and}
    \RErr p  
& = \sup_{\ket{\psi} \in \mathscr{S}}
       \Bigl(\mat{\psi \otimes \pt}{\Epi^2}{\psi \otimes \pt} \Bigr)^{\frac{1}{2}} 
  = \frac{\hbar}{ \sqrt{2} \inf_{\ket{\psi} \in \mathscr{S}} \left(\lambda_{\psi}\right) }
\end{align*}
where $\mathscr{S}$ denotes the unit sphere in the system state space.  In view of
Eq.~(\ref{eq:  RetOptCond2}) it then follows
\begin{equation*}
    \inf_{\ket{\psi} \in \mathscr{S}} \left(\lambda_{\psi}\right)
=   \sup_{\ket{\psi} \in \mathscr{S}} \left(\lambda_{\psi}\right)
\end{equation*}
which means that $\lambda_{\psi}$ must be constant.
\end{proof}
We next define the operators
\begin{equation}
\begin{split}
    \hat{c}^{\vphantom{\dagger}}_{\lambda_{\mathrm{i}}} 
& = \frac{1}{\sqrt{2}} \left( \frac{1}{\lambda_{\mathrm{i}}} \Exi - \frac{i \lambda_{\mathrm{i}}}{\hbar} \Epi\right) \\
    \hat{c}^{\dagger}_{\lambda_{\mathrm{i}}} 
& = \frac{1}{\sqrt{2}} \left( \frac{1}{\lambda_{\mathrm{i}}} \Exi + \frac{i \lambda_{\mathrm{i}}}{\hbar} \Epi\right)
\end{split}
\label{eq:  COpDef}
\end{equation}
In the general case we cannot assume the commutation relation of Eq.~(\ref{eq:  RetErrComRel}).
It follows, that $\hat{c}^{\vphantom{\dagger}}_{\lambda_{\mathrm{i}}}$, $\hat{c}^{\dagger}_{\lambda_{\mathrm{i}}}$
are not, in general, ladder operators.  We do, however, have the relationship of 
Eq.~(\ref{eq:  RetOptComExpect}), and this is enough to prove
\begin{lemma}
Given any retrodictively optimal measurement process with
intial apparatus state $\ket{\pt}$ and retrodictive spatial resolution
$\lambda_{\mathrm{i}}$, let $\hat{c}^{\vphantom{\dagger}}_{\lambda_{\mathrm{i}}}$ be the operator
defined by Eq.~(\ref{eq:  COpDef}).  Then
\begin{equation*}
     \hat{c}^{\vphantom{\dagger}}_{\lambda_{\mathrm{i}}} \ket{\psi \otimes \pt}
=    0
\end{equation*}
for every $\ket{\psi}\in \mathscr{H}_{\mathrm{sy}}$.
\label{lem:  cZero}
\end{lemma}
\begin{proof}
Given any normalised system state $\ket{\psi}$, let $\alpha$, $\beta \in \mathbb{R}$ be the real 
and imaginary parts of $\mat{\psi \otimes \pt}{\Exi\, \Epi}{\psi \otimes \pt}$:
\begin{equation}
    \mat{\psi \otimes \pt}{\Exi\, \Epi}{\psi \otimes \pt} = \alpha + i \beta
\label{eq:  RetErrProdExpct}
\end{equation}
We have
\begin{equation*}
    \left( \alpha^2 + \beta^2 \right)^{\frac{1}{2}}
=   \bigl| \mat{\psi \otimes \pt}{\Exi\, \Epi}{\psi \otimes \pt} \bigr|
\le \bigl\| \Exi \ket{\psi \otimes \pt} \bigr\| \; \bigl\| \Epi \ket{\psi \otimes \pt} \bigr\|
=   \frac{\hbar}{2}
\end{equation*}
where
\begin{equation*}
\begin{split}
    \bigl\| \Exi \ket{\psi \otimes \pt} \bigr\| 
& = \Bigl( \mat{\psi \otimes \pt}{\Exi^2}{\psi \otimes \pt} \Bigr)^{\frac{1}{2}}
  = \frac{\lambda_{\mathrm{i}}}{\sqrt{2}} \\
    \bigl\| \Epi \ket{\psi \otimes \pt} \bigr\| 
& = \Bigl( \mat{\psi \otimes \pt}{\Epi^2}{\psi \otimes \pt} \Bigr)^{\frac{1}{2}}
  = \frac{\hbar}{\sqrt{2} \lambda_{\mathrm{i}}}
\end{split}
\end{equation*}
are the norms of the vectors $\Exi \ket{\psi \otimes \pt}$, $\Epi \ket{\psi \otimes \pt}$.

In view of Eq.~(\ref{eq:  RetOptComExpect}) we also have
\begin{equation*}
  - i \hbar = \mat{\psi \otimes \pt}{\comm{\Exi}{\Epi}}{\psi \otimes \pt} = 2 i \beta
\end{equation*}
Consequently, $\alpha = 0$ and $\beta = - \frac{\hbar}{2}$.  We then have
\begin{equation*}
  \bigl|  \mat{\psi \otimes \pt}{\Exi\, \Epi}{\psi \otimes \pt} \bigr|
= \frac{\hbar}{2}
= \bigl\| \Exi \ket{\psi \otimes \pt} \bigr\| \, \bigl\| \Epi \ket{\psi \otimes \pt} \bigr\| 
\end{equation*}
Now it is generally true, in any Hilbert space, that two vectors $\ket{\Psi_1}$,
$\ket{\Psi_2}$ having the property
\begin{equation*}
   \bigl| \overlap{\Psi_1}{\Psi_2} \bigr| = \bigl\| \ket{\Psi_1} \bigr\| \; \bigl\| \ket{\Psi_2} \bigr\|
\end{equation*}
must be parallel.  Hence
\begin{equation*}
   \Epi \ket{\psi \otimes \pt} = \gamma \, \Exi \ket{\psi \otimes \pt}
\end{equation*}
for some $\gamma \in \mathbb{C}$.  Inserting this result into Eq.~(\ref{eq:  RetErrProdExpct})
we find
\begin{equation*}
  \gamma = - \frac{ i \hbar}{\lambda_{\mathrm{i}}^2}
\end{equation*}
The claim follows.
\end{proof}
Now let 
\begin{equation}
    \rho \left( \Mxf, \Mpf \right)
=   \int d \xf \, d\yfv{1} \dots d\yfv{n} \,  
       \bigl| \overlap{\xf, \Mxf,\Mpf, \yfv{1}, \dots, \yfv{n}}{\psi \otimes \pt} \bigr|^2
\label{eq:  RhoDef}
\end{equation}
be the probability distribution for the final pointer positions.  In this expression
$\ket{\xf, \Mxf,\Mpf, \yfv{1}, \dots, \yfv{n}}$ is the simultaneous eigenvector of the Heisenberg
picture operators $\xfOp$, $\MxfOp$, $\MpfOp$, $\yfOp{j}$, with eigenvalues
$\xf$, $\Mxf$, $\Mpf$, $\yfv{j}$.  We continue to employ the notation and terminology
of ref.~\cite{self2}.  Thus, $\xfOp$ is the final system position operator, $\MxfOp$ and
$\MpfOp$ are the final pointer position operators, and the $\yfOp{j}$ represent the additional, internal degrees
of freedom characterising the apparatus.

Let $\ket{\coh{x}{p}{\lambda_{\mathrm{i}}}} \in \mathscr{H}_{\mathrm{sy}}$ be the state with
wave function
\begin{equation}
   \overlap{x'}{\coh{x}{p}{\lambda_{\mathrm{i}}}}
=  \left( \frac{1}{\pi \lambda_{\mathrm{i}}^2} \right)^{\frac{1}{4}} 
   \exp \left[ - \tfrac{1}{2 \lambda_{\mathrm{i}}^2} (x'-x)^2 + \tfrac{i}{\hbar} p x' - \tfrac{i}{2 \hbar} p x \right]
\label{eq:  CohSteDef}
\end{equation}
and let 
\begin{equation}
   Q_{\lambda_{\mathrm{i}}} (x, p)
=  \frac{1}{h} \, \left| \boverlap{  \coh{x}{p}{\lambda_{\mathrm{i}}} }{ \psi } \right|^2
\label{eq:  HusFuncDef}
\end{equation}
be the initial system state Husimi function~\cite{Husimi,Reviews}.  We want to 
show
\begin{equation*}
  \rho \left( \Mxf, \Mpf \right)  =  Q_{\lambda_{\mathrm{i}}} \left( \Mxf, \Mpf \right)
\end{equation*}
for almost all $\Mxf$, $\Mpf$ whenever the measurement is retrodictively optimal at spatial
resolution $\lambda_{\mathrm{i}}$ (``almost all'' being defined relative to ordinary Lebesgue measure
on the plane).  Our strategy will be to begin by showing that the two functions have the same
moments:
\begin{equation*}
    \int d \Mxf d\Mpf \, \Mxf^n \Mpf^m \, \rho \left( \Mxf, \Mpf \right)
=   \int d \Mxf d\Mpf \, \Mxf^n \Mpf^m \,  Q_{\lambda_{\mathrm{i}}} \left( \Mxf, \Mpf \right)
\end{equation*}
for every pair of non-negative integers $n$, $m$.  Unfortunately we then face the difficulty,
that although $\rho$ and $Q_{\lambda_{\mathrm{i}}}$ are 
always defined, whatever the initial state of the system,
the same is not true of their  moments.  This is because $\xiOp$, $\piOp$, $\MxfOp$, $\MpfOp$ are unbounded
operators.  The way in which we will
circumvent the difficulty is, first to prove the result on the assumption that 
$\ket{\psi}$ is in an appropriately chosen dense subspace of $\mathscr{H}_{\mathrm{sy}}$, and then
to use a continuity argument to extend it to the case of arbitrary $\ket{\psi}$.

Let $\hat{a}_{\lambda_\mathrm{i}}^{\vphantom{\dagger}}$,
$\hat{a}_{\lambda_\mathrm{i}}^{\dagger}$ be the ladder operators
\begin{equation}
\begin{split}
    \hat{a}_{\lambda_\mathrm{i}}^{\vphantom{\dagger}}
& = \frac{1}{\sqrt{2}} \left( \frac{1}{\lambda_\mathrm{i}} \xiOp + \frac{\lambda_\mathrm{i}}{\hbar} \piOp \right) \\
    \hat{a}_{\lambda_\mathrm{i}}^{\dagger}
& = \frac{1}{\sqrt{2}} \left( \frac{1}{\lambda_\mathrm{i}} \xiOp - \frac{\lambda_\mathrm{i}}{\hbar} \piOp \right)
\end{split}
\label{eq:  aOpDef}
\end{equation}
and define number states $\ket{n}_{\lambda_\mathrm{i}} \in \mathscr{H}_{\mathrm{sy}}$ 
in the usual way, by the requirements
\begin{equation*}
    \hat{a}_{\lambda_\mathrm{i}}^{\vphantom{\dagger}} \ket{0}_{\lambda_\mathrm{i}}
=   0 \hspace{0.5 in}
    \suboverlapB{0}{0}{\lambda_{\mathrm{i}}}{5}_{\lambda_{\mathrm{i}}}
=   1 \hspace{0.5 in}
    \ket{n}_{\lambda_\mathrm{i}} 
= \frac{1}{\sqrt{n !}} \, \hat{a}_{\lambda_\mathrm{i}}^{\dagger \; n} \, \ket{0}_{\lambda_\mathrm{i}}
\end{equation*}
(with a slight abuse of notation we sometimes regard the operators $\xiOp$ and $\piOp$
as acting on $\mathscr{H}_{\mathrm{sy}}$, and sometimes as acting on 
$\mathscr{H}_{\mathrm{sy}} \otimes \mathscr{H}_{\mathrm{ap}}$).  We then define 
$\mathscr{F}_{\lambda_{\mathrm{i}}}$ to be the dense subspace of $  \mathscr{H}_{\mathrm{sy}}$
consisting of all \emph{finite} linear combinations of the vectors $\ket{n}_{\lambda_{\mathrm{i}}}$.

It is easily seen that
$\mathscr{F}_{\lambda_{\mathrm{i}}}$
is in the domain of definition of every polynomial $f(\xiOp, \piOp)$.  In particular,
the integral
\begin{equation*}
  \int d x  dp \, 
          x^n p^m Q_{\lambda_{\mathrm{i}}} \left(x, p \right)
\end{equation*}
is defined and finite for all $n$, $m$ whenever $Q_{\lambda_{\mathrm{i}}}$ is the Husimi
function corresponding to a state in $\mathscr{F}_{\lambda_{\mathrm{i}}}$.

Now define the operators
\begin{equation*}
\begin{split}
    \hat{b}_{\lambda_{\mathrm{i}}}^{\vphantom{\dagger}}
& = \frac{1}{\sqrt{2}} 
    \left( \frac{1}{\lambda_{\mathrm{i}}} \MxfOp + \frac{i \lambda_{\mathrm{i}}}{\hbar} \MpfOp \right) \\
    \hat{b}_{\lambda_{\mathrm{i}}}^{\dagger}
& = \frac{1}{\sqrt{2}} 
    \left( \frac{1}{\lambda_{\mathrm{i}}} \MxfOp - \frac{i \lambda_{\mathrm{i}}}{\hbar} \MpfOp \right) 
\end{split}
\end{equation*}
These operators commute, and so they are certainly not ladder operators.
We have
\begin{equation}
    \hat{b}_{\lambda_{\mathrm{i}}}^{\dagger}
=   \hat{a}_{\lambda_{\mathrm{i}}}^{\dagger} + \hat{c}_{\lambda_{\mathrm{i}}}^{\vphantom{\dagger}}
\end{equation}
where $\hat{c}_{\lambda_{\mathrm{i}}}^{\vphantom{\dagger}}$ and $\hat{a}_{\lambda_{\mathrm{i}}}^{\dagger}$
are the operators defined in Eqs.~(\ref{eq:  COpDef}) and~(\ref{eq:  aOpDef}) respectively.
Let $\ket{\psi}$ be any vector $\in \mathscr{F}_{\lambda_{\mathrm{i}}}$.  Then
$\ket{\psi \otimes \pt}$ is in the domain of $\hat{a}_{\lambda_{\mathrm{i}}}^{\dagger}$.
It is also in the domain  of  
$\hat{c}_{\lambda_{\mathrm{i}}}^{\vphantom{\dagger}}$ (the definition of a retrodictively 
optimal process tacitly assumes that $\ket{\psi \otimes \pt}$ is in the domain of
$\Exi$, $\Epi$, and therefore in the domain of $\hat{c}_{\lambda_{\mathrm{i}}}^{\vphantom{\dagger}}$,
for all $\ket{\psi}$).
It is consequently in the domain  of 
$\hat{b}_{\lambda_{\mathrm{i}}}^{\dagger}$.  Moreover, in view of
Lemma~\ref{lem:  cZero},
\begin{equation*}
    \hat{b}_{\lambda_{\mathrm{i}}}^{\dagger} \ket{\psi \otimes \pt}
=   \left( \hat{a}_{\lambda_{\mathrm{i}}}^{\dagger} \ket{\psi} \right) \otimes \ket{\pt}
\end{equation*}
where $\hat{a}_{\lambda_{\mathrm{i}}}^{\dagger} \ket{\psi}$ also $\in \mathscr{F}_{\lambda_{\mathrm{i}}}$.
Iterating the argument we conclude that $\ket{\psi \otimes \pt}$ is in the domain of
$\hat{b}_{\lambda_{\mathrm{i}}}^{\dagger \; n}$ and
\begin{equation*}
    \hat{b}_{\lambda_{\mathrm{i}}}^{\dagger \; n} \ket{\psi \otimes \pt}
=   \left( \hat{a}_{\lambda_{\mathrm{i}}}^{\dagger \; n} \ket{\psi} \right) \otimes \ket{\pt}
\end{equation*}
for every non-negative integer $n$.  Taking adjoints gives
\begin{equation*}
   \bra{\psi \otimes \pt} \hat{b}_{\lambda_{\mathrm{i}}}^{\vphantom{\dagger} \; m} 
=  \left( \bra{\psi} \hat{a}_{\lambda_{\mathrm{i}}}^{\vphantom{\dagger} \; m} \right) \otimes \bra{\pt}
\end{equation*}
for all $m$.  Consequently,
\begin{equation*}
  \bmat{\psi \otimes \pt}{ \hat{b}_{\lambda_{\mathrm{i}}}^{\vphantom{\dagger} \; m}
                          \hat{b}_{\lambda_{\mathrm{i}}}^{\dagger \; n}
                       }{ \psi \otimes \pt
                        } 
=  \bmat{\psi}{  \hat{a}_{\lambda_{\mathrm{i}}}^{\vphantom{\dagger} \; m}
                \hat{a}_{\lambda_{\mathrm{i}}}^{\dagger \; n}
            }{  \psi
             }
\end{equation*}
Now
\begin{equation*}
  \bmat{\psi \otimes \pt}{ \hat{b}_{\lambda_{\mathrm{i}}}^{\vphantom{\dagger} \; m}
                          \hat{b}_{\lambda_{\mathrm{i}}}^{\dagger \; n}
                       }{ \psi \otimes \pt
                        } 
= \int d\Mxf d\Mpf \,  z_{\lambda_{\mathrm{i}}}^m z_{\lambda_{\mathrm{i}}}^{* \; n} \, \rho \left( \Mxf, \Mpf \right)
\end{equation*}
where $\rho$ is the distribution of final pointer positions, as defined in Eq.~(\ref{eq:  RhoDef}),
and $z_{\lambda_{\mathrm{i}}}$ is the complex coordinate
\begin{equation}
   z_{\lambda_{\mathrm{i}}} 
= \frac{1}{\sqrt{2}} \left( \frac{1}{\lambda_{\mathrm{i}}} \Mxf + \frac{i \lambda_{\mathrm{i}}}{\hbar} \Mpf\right)
\label{eq:  ZedDef}
\end{equation}
Also~\cite{Reviews}
\begin{equation}
   \bmat{\psi}{  \hat{a}_{\lambda_{\mathrm{i}}}^{\vphantom{\dagger} \; m}
                \hat{a}_{\lambda_{\mathrm{i}}}^{\dagger \; n}
            }{  \psi
             }
=  \int d\Mxf d\Mpf \,  
    z_{\lambda_{\mathrm{i}}}^m z_{\lambda_{\mathrm{i}}}^{* \; n} \, Q_{\lambda_{\mathrm{i}}} \left( \Mxf, \Mpf \right) 
\label{eq:  HusMoms}
\end{equation}
where $Q_{\lambda_{\mathrm{i}}}$ is the initial system state Husimi function, as defined in
Eq.~(\ref{eq:  HusFuncDef}).  Therefore
\begin{equation*}
   \int d\Mxf d\Mpf \,  z_{\lambda_{\mathrm{i}}}^m z_{\lambda_{\mathrm{i}}}^{* \; n} \, \rho \left( \Mxf, \Mpf \right)
=  \int d\Mxf d\Mpf \,  
    z_{\lambda_{\mathrm{i}}}^m z_{\lambda_{\mathrm{i}}}^{* \; n} \, Q_{\lambda_{\mathrm{i}}} \left( \Mxf, \Mpf \right)
\end{equation*}
for all $n$, $m$.  It follows that
\begin{equation*}
   \int d\Mxf d\Mpf \,  f ( z_{\lambda_{\mathrm{i}}}, z_{\lambda_{\mathrm{i}}}^{*} ) \,
                     \rho \left( \Mxf, \Mpf \right)
=  \int d\Mxf d\Mpf \,  
    f ( z_{\lambda_{\mathrm{i}}}, z_{\lambda_{\mathrm{i}}}^{*} ) \, 
    Q_{\lambda_{\mathrm{i}}} \left( \Mxf, \Mpf \right)
\end{equation*}
for every polynomial $f$.  In particular
\begin{equation}
   \int d\Mxf d\Mpf \,  \Mxf^m \Mpf^n \,
                     \rho \left( \Mxf, \Mpf \right)
=  \int d\Mxf d\Mpf \,  
    \Mxf^m \Mpf^n \, 
    Q_{\lambda_{\mathrm{i}}} \left( \Mxf, \Mpf \right)
\label{eq:  Moments}
\end{equation}
for all $m$, $n$.  

At this stage one needs to be careful.  It is tempting to suppose, that two probability 
measures which have the same moments must be equal.  In fact, this inference is not always
justified (see Reed and Simon~\cite{Reed}, vol.~2).  However, it is justified here, as we show in 
the Appendix.  Consequently
\begin{equation}
  \rho \left( \Mxf, \Mpf \right) = Q_{\lambda_{\mathrm{i}}} \left( \Mxf, \Mpf \right)
\label{eq:  EqOfMeas}
\end{equation}
for almost all $\Mxf$, $\Mpf$ whenever the initial system state $\ket{\psi}$ is
in the space $\mathscr{F}_{\lambda_{\mathrm{i}}}$.

It remains for us to show that the distributions are equal in the case of arbitrary
$\ket{\psi} \in \mathscr{H}_{\mathrm{sy}}$.  We will do this by using a continuity 
argument.

Choose
a sequence $\ket{\psi_n} \in \mathscr{F}_{\lambda_{\mathrm{i}}}$ converging 
to $\ket{\psi}$.  Let $Q_{\lambda_{\mathrm{i}}, n}$ be the Husimi function, 
and $\rho_n$ the distribution of measured values corresponding to 
$\ket{\psi_n}$.  Let $Q_{\lambda_{\mathrm{i}}}$ be the Husimi function, 
and $\rho$ the distribution of measured values corresponding to 
$\ket{\psi}$.  

We have, as an immediate consequence of the definition, Eq.~(\ref{eq:  HusFuncDef}),
\begin{equation}
    Q_{\lambda_{\mathrm{i}}} \left( \Mxf, \Mpf \right)
=  \lim_{n \rightarrow \infty} \bigl( Q_{\lambda_{\mathrm{i}}, n} \left( \Mxf, \Mpf \right)   \bigr)
\label{eq:  HusFuncLim}
\end{equation}
for all $\Mxf$, $\Mpf$.  

On the other hand, it is not generally true that $\rho_n$ converges pointwise
to $\rho$.  It does, however, contain a subsequence which converges pointwise almost everywhere.
In fact, let $\mathscr{L}_1$ be the Banach space consisting of all integrable functions on
$\mathbb{R}^2$, with norm
\begin{equation*}
   \left\| f \right\|_1 = \int d \Mxf d \Mpf \left|f\left(\Mxf,\Mpf\right) \right|
\end{equation*}
We have
\begin{align*}
    \left\| \rho - \rho_n \right\|_1
& = \int d\Mxf d\Mpf \,
       \left| \int d \xf d \yfv{1} \dots \yfv{n} \, 
                 \left(   \left| \overlap{ \xf,\Mxf,\Mpf,\yfv{1}, \dots, \yfv{n}
                                        }{ \psi \otimes \pt}
                          \right|^2
                 \right. \right.
\\
& \hspace{1.75 in}
       \left. \left.
                       -  \left|\overlap{  \xf,\Mxf,\Mpf,\yfv{1},\dots,\yfv{n}
                                       }{  \psi_n \otimes \pt
                                       }
                         \right|^2
                 \right)
       \right|
\\
& \le  \bigl\| \ket{\psi \otimes \pt} - \ket{\psi_n \otimes \pt} \bigr\| \ 
       \Bigl( \bigl\|\ket{\psi \otimes \pt} \bigr\| + \bigl\| \ket{\psi \otimes \pt} \bigr\| \Bigr)
\\
& \rightarrow 0
\end{align*}
We see from this that $\rho_n \rightarrow \rho$ in the topology of $\mathscr{L}_1$.  We may therefore
use the Riesz-Fisher theorem (Reed and Simon~\cite{Reed}, vol.~1) to deduce that it contains a 
subsequence $\rho_{n_r}$ such that
\begin{equation*}
  \rho \left( \Mxf, \Mpf \right) = \lim_{r \rightarrow \infty} \bigl( \rho_{n_r} \left(\Mxf, \Mpf\right)  \bigr)
\end{equation*}
for almost all $\Mxf$, $\Mpf$.  In view of this result, Eq.~(\ref{eq:  HusFuncLim}), and the fact that
\begin{equation*}
    \rho_{n_r} \left( \Mxf, \Mpf\right) = \Omega_{\lambda_{\mathrm{i}}, n_r} \left( \Mxf, \Mpf\right)
\end{equation*}
for all $r$ and almost all $\Mxf$, $\Mpf$ we deduce that
\begin{equation*}
  \rho \left( \Mxf, \Mpf\right)  =  \Omega_{\lambda_{\mathrm{i}}} \left( \Mxf, \Mpf\right)
\end{equation*}
for almost all $\Mxf$, $\Mpf$.
\section{Predictively Optimal Measurements}
\label{sec:  PreOpt}
We will say that a simultaneous measurement process of the kind defined in ref.~\cite{self2}
is \emph{predictively optimal} if the product of predictive errors is minimised:
\begin{equation}
    \PErr x \, \PErr p = \frac{\hbar}{2}
\label{eq:  PreErrOptCond}
\end{equation}
In view of the commutation relation
\begin{equation}
   \comm{\Exf}{\Epf} = i \hbar
\label{eq:  PreErrComRel}
\end{equation}
there is no need to impose the condition, that the measurement be predictively unbiased as a 
separate requirement:  it is a consequence of the condition of Eq.~(\ref{eq:  PreErrOptCond}).

Eqs.~(\ref{eq:  PreErrOptCond}) and~(\ref{eq:  PreErrComRel}) together imply
\begin{equation*}
  \mat{\psi \otimes \pt}{\Exf^2}{\psi \otimes \pt} \,
  \mat{\psi \otimes \pt}{\Epf^2}{\psi \otimes \pt}
= \frac{\hbar^2}{4}
\end{equation*}
for every normalised $\ket{\psi} \in \mathscr{H}_{\mathrm{sy}}$.  By an argument which
parallels the proof of Lemma~\ref{lem:  RetRes} we infer that there exists a fixed
number $\lambda_{\mathrm{f}}$ such that
\begin{equation*}
\begin{split}
   \mat{\psi \otimes \pt}{\Exf^2}{\psi \otimes \pt} & = \frac{\lambda_{\mathrm{f}}^2}{2} \\
   \mat{\psi \otimes \pt}{\Epf^2}{\psi \otimes \pt} & = \frac{\hbar^2}{2 \lambda_{\mathrm{f}}^2}
\end{split}
\end{equation*}
for every normalised $\ket{\psi}\in\mathscr{H}_{\mathrm{sy}}$.  It is then straightforward to show that
\begin{equation}
   \hat{d}_{\lambda_{\mathrm{f}}} \ket{\psi \otimes \pt} = 0
\label{eq:  dOpAction}
\end{equation}
for all $\ket{\psi}\in\mathscr{H}_{\mathrm{sy}}$, where $\hat{d}_{\lambda_{\mathrm{f}}}$
is the annihilation operator
\begin{equation*}
   \hat{d}_{\lambda_{\mathrm{f}}}  
=  \frac{1}{\sqrt{2}} \left( \frac{1}{\lambda_{\mathrm{f}}} \Exf + \frac{i\lambda_{\mathrm{f}}}{\hbar} \Epf \right)
\end{equation*}
Since $\Exf$, $\Epf$ are canonically conjugate there exist kets 
$\ket{\Exfv, \Mxf, \Mpf, \yfv{1},\dots , \yfv{n}}_{\epsilon}$ which are simultaneous 
eigenvectors of the operators $\Exf$, $\MxfOp$, $\MpfOp$, $\yfOp{j}$, and which have the
property
\begin{equation}
   \submatB{\Exfv, \Mxf, \Mpf, \yfv{1},\dots ,\yfv{n}}{\Epf}{\Psi}{\epsilon}{4}
=  - i \hbar \frac{\partial}{\partial \Exfv} \; 
    \suboverlapB{\Exfv, \Mxf, \Mpf, \yfv{1},\dots , \yfv{n}}{\Psi}{\epsilon}{4}
\label{eq:  EpsRepDef}
\end{equation}
for all $\ket{\Psi} \in \mathscr{H}_{\mathrm{sy}} \otimes \mathscr{H}_{\mathrm{ap}}$.
In view of Eq.~(\ref{eq:  dOpAction}) we then have
\begin{equation*}
    \left( \frac{1}{\lambda_{\mathrm{f}}} \Exfv + \lambda_{\mathrm{f}} \frac{\partial}{\partial \Exfv}
    \right)
    \suboverlapB{\Exfv, \Mxf, \Mpf, \yfv{1}, \dots, \yfv{n}}{\psi \otimes \pt}{\epsilon}{4}
=   0
\end{equation*}
for all $\ket{\psi} \in \mathscr{H}_{\mathrm{sy}}$.  Solving this equation
we find
\begin{align}
&   \suboverlapB{\Exfv, \Mxf, \Mpf, \yfv{1},\dots , \yfv{n}}{\psi \otimes \pt}{\epsilon}{4}
\notag
\\
& \hspace{0.25 in}
 =    \left( \frac{1}{\pi \lambda_{\mathrm{f}}^2} \right)^{\frac{1}{4}}
    \exp\left[ - \tfrac{1}{2 \lambda_{\mathrm{f}}^2} \Exfv^2 \right] 
    \Phi \left( \Mxf, \Mpf, \yfv{1}, \dots , \yfv{n}\right)
\label{eq:  eRepFinWveFnc}
\end{align}
where $\Phi$ is an arbitrary normalised function.

There also exist kets $\ket{\xf,\Mxf,\Mpf,\yfv{1}, \dots , \yfv{n}}_{x}$ 
which are simultaneous eigenvectors of the
operators $\xfOp$, $\MxfOp$, $\MpfOp$, $\yfOp{j}$ with the property
\begin{equation}
    \submatB{\xf,\Mxf,\Mpf,\yfv{1}, \dots , \yfv{n}}{\pfOp}{\Psi}{x}{4}
=   - i \hbar \frac{\partial}{\partial \xf} \; 
       \suboverlapB{\xf,\Mxf,\Mpf,\yfv{1}, \dots , \yfv{n}}{\Psi}{x}{4}
\label{eq:  xRepDef}
\end{equation}
for all $\ket{\Psi} \in \mathscr{H}_{\mathrm{sy}} \otimes \mathscr{H}_{\mathrm{ap}}$.  
In view of the defining relation
$\Exf = \MxfOp - \xfOp$ we must have
\begin{align}
&   \ket{\xf,\Mxf,\Mpf,\yfv{1}, \dots , \yfv{n}}_x 
\notag
\\
& \hspace{0.5 in}
=   e^{- i \chi \left( \xf,\Mxf,\Mpf,\yfv{1}, \dots , \yfv{n}\right)}
      \ket{\Mxf - \xf, \Mxf, \Mpf, \yfv{1}, \dots , \yfv{n}}_{\epsilon}
\label{eq:  xRepDefB}
\end{align}
where $e^{- i \chi \left( \xf,\Mxf,\Mpf,\yfv{1}, \dots , \yfv{n}\right)}$ is a phase.  In view
of Eqs.~(\ref{eq:  EpsRepDef}) and~(\ref{eq:  xRepDef}) we must then have
\begin{align*}
&  \submatB{\xf,\Mxf,\Mpf,\yfv{1}, \dots , \yfv{n}}{\pfOp}{\Psi}{x}{4}
\\
& \hspace{0.5 in}
=  - i \hbar \frac{\partial}{\partial \xf}
   \left( e^{i \chi \left( \xf,\Mxf,\Mpf,\yfv{1}, \dots , \yfv{n}\right)} 
   \suboverlapB{\Mxf - \xf,\Mxf,\Mpf,\yfv{1}, \dots , \yfv{n}}{\Psi}{\epsilon}{4}
   \right)
\end{align*}
and
\begin{align*}
&    \submatB{\xf,\Mxf,\Mpf,\yfv{1}, \dots , \yfv{n}}{\pfOp}{\Psi}{x}{4}
\\
&  \hspace{0.25 in}
  =  e^{i \chi \left( \xf,\Mxf,\Mpf,\yfv{1}, \dots , \yfv{n}\right)}
     \submatB{\Mxf - \xf,\Mxf,\Mpf,\yfv{1}, \dots , \yfv{n}}{\MpfOp-\Epf}{\Psi}{\epsilon}{4}
\\
& \hspace{0.25 in}
  =  e^{i \chi \left( \xf,\Mxf,\Mpf,\yfv{1}, \dots , \yfv{n}\right)}
\\
& \hspace{0.75 in}
     \times
     \left( \Mpf + i \hbar \frac{\partial}{\partial \Exf} \right)
     \left. \suboverlapB{      \Exf,\Mxf,\Mpf,\yfv{1}, \dots , \yfv{n}
                       }{      \Psi
                       }{      \epsilon}{4} 
     \right|_{\Exf = \Mxf - \xf}
\end{align*}
for all $\ket{\Psi} \in \mathscr{H}_{\mathrm{sy}} \otimes \mathscr{H}_{\mathrm{ap}}$.
Hence
\begin{equation*}
   \hbar \frac{\partial }{\partial \xf} 
   \chi\left( \xf,\Mxf,\Mpf,\yfv{1}, \dots , \yfv{n}\right) = \Mpf
\end{equation*}
which implies
\begin{equation*}
  \chi\left( \xf,\Mxf,\Mpf,\yfv{1}, \dots , \yfv{n}\right) 
= \frac{1}{\hbar} \Mpf \xf + \chi_0 \left( \Mxf,\Mpf,\yfv{1}, \dots , \yfv{n}\right)
\end{equation*}
where $\chi_0$ is an arbitrary function.  
Using this result and Eq.~(\ref{eq:  xRepDefB}) in Eq.~(\ref{eq:  eRepFinWveFnc}) we 
deduce, that the final state wave function can be written
\begin{align*}
&  \suboverlapB{\xf, \Mxf, \Mpf, \yfv{1}, \dots , \yfv{n}}{\psi \otimes \pt}{x}{4}
\\
& \hspace{0.25 in}
=   \left( \frac{1}{\pi \lambda_{\mathrm{f}}^2} \right)^{\frac{1}{4}}
    \exp \bigl[ - \tfrac{1}{2 \lambda_{\mathrm{f}}^2 } \left( \Mxf - \xf\right)^2 
                + \tfrac{i}{\hbar} \Mpf \xf 
                + i \chi_0 \left( \Mxf,\Mpf,\yfv{1}, \dots , \yfv{n}\right)
         \bigr]
\\
& \hspace{ 2.5 in}
    \times
    \Phi \left( \Mxf,\Mpf,\yfv{1}, \dots , \yfv{n}\right)
\end{align*}
In terms of the state $\ket{\coh{\Mxf}{\Mpf}{\lambda_{\mathrm{f}}}}$ 
defined in Eq.~(\ref{eq:  CohSteDef}) this becomes
\begin{equation*}
    \suboverlapB{\xf, \Mxf, \Mpf, \yfv{1}, \dots , \yfv{n}}{\psi \otimes \pt}{x}{4}
=   \boverlap{\xf}{\coh{\Mxf}{\Mpf}{\lambda_{\mathrm{f}}}} \, 
     \Phi' \left( \Mxf,\Mpf,\yfv{1}, \dots , \yfv{n}\right)
\end{equation*}
where
\begin{align*}
&  \Phi' \left( \Mxf,\Mpf,\yfv{1}, \dots , \yfv{n}\right) 
\\
& \hspace{0.25 in}
=   \exp \bigl[   i \chi_0 \left( \Mxf,\Mpf,\yfv{1}, \dots , \yfv{n}\right) 
               + \tfrac{i}{2 \hbar} \Mpf \Mxf 
         \bigr] \,
   \Phi \left( \Mxf,\Mpf,\yfv{1}, \dots , \yfv{n}\right)
\end{align*}
The distribution of measured values $\rho \left( \Mxf, \Mpf \right)$ 
can be written in terms of $\Phi'$: 
\begin{equation*}
  \rho \left( \Mxf, \Mpf \right) 
= \int d \yfv{1} \dots \yfv{n} \, 
     \left|\Phi' \left( \Mxf, \Mpf, \yfv{1}, \dots , \yfv{n} \right) \right|^2
\end{equation*}
Suppose, now, that the pointer positions are found to be in the
region $\mathscr{R} \subseteq \mathbb{R}^2$. Let $\hat{\rho}_{\mathrm{sy}}$ be 
the reduced density matrix describing the state of the system immediately
afterwards.  Then
\begin{align*}
    \mat{x_{\mathrm{f}1}}{\hat{\rho}_{\mathrm{sy}}}{x_{\mathrm{f}2}}
& = \frac{1}{p_{\mathscr{R}}}
    \int_{\mathscr{R} \times \mathbb{R}^n} d\Mxf d\Mpf d\yfv{1} \dots d \yfv{n} \,
       \left| \Phi' \left( \Mxf , \Mpf, \yfv{1}, \dots , \yfv{n} \right) \right|^2
\\
& \hspace{1.75 in}
    \times
       \boverlap{x_{\mathrm{f}1}}{\coh{\Mxf}{\Mpf}{\lambda_{\mathrm{f}}} }
       \boverlap{  \coh{\Mxf}{\Mpf}{\lambda_{\mathrm{f}}} }{x_{\mathrm{f}2}}
\end{align*}
where $p_{\mathscr{R}}$ is the probability of 
finding $\left( \Mxf, \Mpf \right) \in \mathscr{R}$:
\begin{equation*}
  p_{\mathscr{R}} = \int_{\mathscr{R}} d \Mxf d\Mpf \, \rho \left( \Mxf, \Mpf\right)
\end{equation*}
Hence
\begin{equation*}
  \hat{\rho}_{\mathrm{sy}}
=  \frac{1}{p_{\mathscr{R}}} \int_{\mathscr{R}} d \Mxf d\Mpf \, 
       \rho \left( \Mxf , \Mpf \right)
       \bket{  \coh{\Mxf}{\Mpf}{\lambda_{\mathrm{f}} }  }
       \bbra{  \coh{\Mxf}{\Mpf}{ \lambda_{\mathrm{f}} }  }     
\end{equation*}
On the other hand
\begin{equation*}
  \hat{\rho}_{\mathrm{sy}}
=  \int d \Mxf d\Mpf \, 
       P_{\lambda_{\mathrm{f}}} \left( \Mxf , \Mpf \right)
       \bket{  \coh{\Mxf}{\Mpf}{\lambda_{\mathrm{f}} }  }
       \bbra{  \coh{\Mxf}{\Mpf}{ \lambda_{\mathrm{f}} }  }   
\end{equation*}
where $P_{\lambda_{\mathrm{f}}}$ is the anti-Husimi function 
(or $P$-function)~\cite{Reviews,PRep}
describing the final state of the system.  Comparing these expressions we see
\begin{equation}
  P_{\lambda_{\mathrm{f}}} \left( \Mxf , \Mpf \right)
= \begin{cases}
       \frac{1}{p_{\mathscr{R}}} \rho \left( \Mxf , \Mpf \right) 
              \qquad & \text{if} \left( \Mxf , \Mpf \right) \in \mathscr{R} \\
       0 \qquad & \text{otherwise}
  \end{cases}
\label{eq:  PreOptOut}
\end{equation}
If $\mathscr{R}$ is a sufficiently small region centred on the point 
$\left( \Mxf , \Mpf \right)$ the system is approximately in the state
$\bket{  \coh{\Mxf}{\Mpf}{\lambda_{\mathrm{f}} }  }$ after the measurement:
\begin{equation*}
   \hat{\rho}_{\mathrm{sy}}  \approx 
       \bket{  \coh{\Mxf}{\Mpf}{\lambda_{\mathrm{f}} }  }
       \bbra{  \coh{\Mxf}{\Mpf}{ \lambda_{\mathrm{f}} }  }
\end{equation*}

Eq.~(\ref{eq:  PreOptOut}) shows that the effect of a predictively
optimal measurement process is to leave the system in a state
for which $P_{\lambda_{\mathrm{f}}}$ is a probability density function.
Such states 
are, of course, exceptional.  In many cases, $P_{\lambda_{\mathrm{f}}}$
is not even defined as a tempered distribution~\cite{Reviews}.  
\section{The Interpretation of the Husimi Function}
\label{sec:  Interp}
The result proved in Section~\ref{sec:  RetroOpt} shows that there is a certain  
analogy between the Husimi function and the $x$-space probability
density function $\left| \overlap{x}{\psi}\right|^2$.  To see this let us
examine just what is meant by the statement, that $\left| \overlap{x}{\psi}\right|^2 \delta x$
represents the probability of finding the position to lie in the interval
$\left( x, x+ \delta x\right)$.

Consider a measurement of $x$ only.  For the sake of simplicity suppose that the
measuring apparatus has only one degree of freedom, corresponding to the single
pointer observable $\MxOp$ (the argument which follows does not depend on this
assumption, however).  Let $\ket{\psi}$ and $\ket{\pt}$ be the initial states
of the system and apparatus respectively, and let $\hat{U}$ be the unitary 
evolution operator describing the measurement interaction.  Let
$\xiOp = \xOp$ and
$\MxfOp = \hat{U}^{\dagger} \Mxf \hat{U}$ be the Heisenberg
picture operators describing the initial position of the system 
and final position of the pointer respectively.  Let
$\Exi = \MxfOp - \xiOp$ be the retrodictive error operator.

The final state wave function
can be written (in the Schr\"{o}dinger picture)
\begin{equation*}
    \bmat{x,\Mx}{\hat{U}}{\psi \otimes \pt}
=   \int dx' \, K \left( x, \Mx; x'\right) \overlap{x'}{\psi}
\end{equation*}
for some kernel $K$.  The probability distribution describing the result 
of the measurement then takes the form
\begin{equation}
    \rho \left( \Mx \right) 
=   \int dx \, \left| \int dx' \, K \left( x, \Mx; x'\right) \overlap{x'}{\psi} \right|^2
\label{eq:  RhoXOnly}
\end{equation}
After a certain amount of algebra one also finds
\begin{equation}
    \mat{\psi \otimes \pt}{\Exi^2}{\psi \otimes \pt}
=   \int dx d\Mx \, 
       \left| \int dx' \left( \Mx - x' \right) K\left(x,\Mx; x'\right) \overlap{x'}{\psi} \right|^2
\label{eq:  RetErrXOnly}
\end{equation}
Suppose that $\RErr x = 0$.  Then we see from Eq.~(\ref{eq:  RetErrXOnly}) that 
$K$ must take the form
\begin{equation*}
  K \left(x,\Mx;x'\right) = f\left(x,\Mx\right) \delta\left( \Mx - x' \right)
\end{equation*}
for some function $f$.  The unitarity of $\hat{U}$ means that $f$ must satisfy
\begin{equation*}
   \int dx \, \left| f\left(x,\Mx\right) \right|^2 = 1
\end{equation*}
Using these results in Eq.~(\ref{eq:  RhoXOnly}) we find
\begin{equation*}
   \rho \left( \Mx \right) = \left| \overlap{\Mx}{\psi} \right|^2
\end{equation*}
whenever the measurement is perfectly accurate for the purposes of retrodiction.

Suppose, on the other hand, that $\RErr x > 0$.  Then $\rho\left(\Mx\right)$
will not  generally coincide with the function 
$\left| \overlap{\Mx}{\psi} \right|^2$.  If $\RErr x$ is small compared with the 
de Broglie wavelength, then we see from Eqs.~(\ref{eq:  RhoXOnly})
and~(\ref{eq:  RetErrXOnly}) that 
$\rho \left( \Mx \right) \approx \left| \overlap{\Mx}{\psi} \right|^2$.  
Otherwise, 
we do not expect the two functions even to be approximately equal.

Although one may possibly approach, one does not expect actually to achieve
the limit of perfect accuracy.  It follows, that one does not expect the function 
$\left| \overlap{\Mx}{\psi} \right|^2$ to describe the outcome of any
practically realisable measurement of position.  

This being so what, exactly, is the significance of
the function $\left| \overlap{\Mx}{\psi} \right|^2$?  In the first place, it serves
as a standard of comparison, against which the outcome of experimentally realisable
measurements can be judged:  in the sense, that the better the measurement, the more closely
does the function $\left| \overlap{\Mx}{\psi} \right|^2$ approximate 
the distribution of actual results.  

In the second place, we see from Eq.~(\ref{eq:  RhoXOnly}) that the outcome 
of  a real measurement of position depends, not only on the state of the system,
\emph{via} the function $\overlap{x'}{\psi}$, but also on the details
of the measurement process, \emph{via} the function $K\left(x,\Mx; x'\right)$.
In the limit of perfect retrodictive accuracy, however, the dependence on the
apparatus (as represented by the kernel $K$) disappears, 
and the distribution of results is determined solely by the state
of the system (as represented by the vector $\ket{\psi}$).   
$\left| \overlap{\Mx}{\psi} \right|^2$ does, so to speak, represent the \emph{intrinsic}
distribution of position, independent of any properties specific to the 
particular measuring instrument employed. 
In a real measurement, by contrast, the outcome is (in a manner of speaking) 
contaminated by instrumental contributions, which one may try to reduce, but can never
entirely eliminate.

One typically regards the function $\left| \overlap{\Mx}{\psi} \right|^2$ simply,
and without qualification, as \emph{the} $x$-space probability distribution.
It owes this canonical status to the two features just mentioned.
The result proved in  Section~\ref{sec:  RetroOpt} shows that the Husimi 
function  has analogous features.
It describes the outcome of 
those measurements which are retrodictively optimal, or ``best''.  It is otherwise
independent of the details of the particular process considered.  It might therefore
be regarded as the canonical probability distribution for position and momentum.

In classical mechanics one has the concept of the ``actual'' 
distribution describing an ensemble of identically prepared systems.  Quantum
mechanics contains no precise analogue for this concept (unless one adopts a 
``hidden-variables'' interpretation~\cite{HiddenVariables}).  Nevertheless, the result proved 
in Section~\ref{sec:  RetroOpt} shows that there are certain resemblances between the
Husimi function and the classical distribution.  The Husimi function is clearly not the same
as the classical distribution.  However, one might reasonably argue that it 
is the closest that quantum mechanics allows us to get to the concept
of a ``real'' or ``objective'' phase space probability distribution.
\appendix
\section*{Appendix.  Proof of Equation~(\ref{eq:  EqOfMeas})}
\label{app:  EqOfMeas}
Rather than working in terms of the functions $\rho$, $Q_{\lambda_{\mathrm{i}}}$ 
it will be convenient, instead, to work in terms
of the measures
\begin{equation*}
\begin{split}
    d \mu_{\rho} & = \rho \left( \Mxf, \Mpf \right) d \Mxf d \Mpf \\
    d \mu_{Q}    & = Q_{\lambda_{\mathrm{i}}} \left( \Mxf, \Mpf \right) d \Mxf d \Mpf
\end{split}
\end{equation*}
We have from Eqs.~(\ref{eq:  HusMoms}) and~(\ref{eq:  Moments})
\begin{equation}
  \int d \mu_{\rho} \, \left|z_{\lambda_{\mathrm{i}}} \right|^{ 2 n}
= \int d \mu_{Q} \, \left|z_{\lambda_{\mathrm{i}}} \right|^{2 n}
= \bmat{\psi}{\hat{a}_{\lambda_{\mathrm{i}}}^{\vphantom{\dagger} n} \hat{a}_{\lambda_{\mathrm{i}}}^{\dagger \; n}}{\psi}
\label{eq:  EvenMoms}
\end{equation}
where $z_{\lambda_{\mathrm{i}}}$ is the complex co-ordinate defined in
Eq.~(\ref{eq:  ZedDef}).  Our strategy will be, first to establish a bound on the
rate at which these quantities grow with increasing $n$, and then to use this to show
that the measures $\mu_{\rho}$, $\mu_Q$ have the same Fourier transform.

$\ket{\psi}$ is in the subspace $\mathscr{F}_{\lambda_{\mathrm{i}}}$.  It 
can therefore be written
\begin{equation*}
   \ket{\psi} = \sum_{r=0}^{l} c_r \ket{r}_{\lambda_{\mathrm{i}}}
\end{equation*}
for some  integer $l$.  Hence
\begin{equation*}
               \bmat{  \psi
                   }{  \hat{a}_{\lambda_{\mathrm{i}}}^{\vphantom{\dagger} n} \hat{a}_{\lambda_{\mathrm{i}}}^{\dagger \; n}
                   }{\psi
                   }
 =   \sum_{r = 0}^{l} \frac{(n+r)!}{r!}  \, \left|c_{r}\right|^2 
\le    \frac{(n+l)!}{l!}
\end{equation*}
Let $\mu$ stand for either of the measures $\mu_{\rho}$, $\mu_Q$.
In view of the inequality just proved, Eq.~(\ref{eq:  EvenMoms}) and the fact
\begin{equation*}
  \left|z_{\lambda_{\mathrm{i}}} \right|^{2 n + 1} 
\le  \frac{1}{2} \left( \left|z_{\lambda_{\mathrm{i}}} \right|^{2n} +\left|z_{\lambda_{\mathrm{i}}} \right|^{2n+2} \right)
\end{equation*}
we have 
\begin{equation*}
    \int d \mu \, \left|z_{\lambda_{\mathrm{i}}} \right|^{n}  
\le \frac{\Gamma \left( \frac{1}{2} n +l+\frac{3}{2}\right)}{\Gamma(l+1)}
\end{equation*}
for every non-negative integer $n$.  Hence
\begin{equation*}
   \sum_{n=0}^{\infty} \frac{1}{n!} 
     \int d \mu \, \left| \beta z_{\lambda_{\mathrm{i}}}^{\vphantom{*}} + \gamma z_{\lambda_{\mathrm{i}}}^{*}\right|^n
<  \infty
\end{equation*}
for all $\beta$, $\gamma \in \mathbb{C}$.  It follows that the functions
$e^{| \beta z_{\lambda_{\mathrm{i}}}^{\vphantom{*}} + \gamma z_{\lambda_{\mathrm{i}}}^{*}|}$
and
$e^{ \beta z_{\lambda_{\mathrm{i}}}^{\vphantom{*}} + \gamma z_{\lambda_{\mathrm{i}}}^{*}}$
are $\mu$-integrable.  We may therefore use Lebesgue's dominated convergence theorem
(Reed and Simon~\cite{Reed}, vol.~1) to infer
\begin{align*}
    \int d \mu_{\rho} \, 
     \exp\left[ \beta z_{\lambda_{\mathrm{i}}}^{\vphantom{*}} + \gamma z_{\lambda_{\mathrm{i}}}^{*}\right]
& = \lim_{N\rightarrow \infty} 
    \left( \sum_{n=0}^{N} \frac{1}{n!} \int d \mu_{\rho} \, 
          \left( \beta z_{\lambda_{\mathrm{i}}}^{\vphantom{*}} + \gamma z_{\lambda_{\mathrm{i}}}^{*}  \right)^n
    \right)
\\
& = \lim_{N\rightarrow \infty} 
    \left( \sum_{n=0}^{N} \frac{1}{n!} \int d \mu_{Q} \, 
          \left( \beta z_{\lambda_{\mathrm{i}}}^{\vphantom{*}} + \gamma z_{\lambda_{\mathrm{i}}}^{*}  \right)^n
    \right)
\\
& = \int d \mu_{Q} \, 
     \exp\left[ \beta z_{\lambda_{\mathrm{i}}}^{\vphantom{*}} + \gamma z_{\lambda_{\mathrm{i}}}^{*}\right]
\end{align*}
for all $\beta$, $\gamma \in \mathbb{C}$.  Consequently
\begin{equation*}
    \int d \mu_{\rho} \, 
     \exp\left[ i\left( k_{\mathrm{X}} \Mxf + k_{\mathrm{P}} \Mpf \right) \right]
=  \int d \mu_{Q} \, 
     \exp\left[ i\left( k_{\mathrm{X}} \Mxf + k_{\mathrm{P}} \Mpf \right) \right]
\end{equation*}
for all $k_{\mathrm{X}}$, $k_{\mathrm{P}} \in\mathbb{R}$.  Inverting the Fourier transforms we deduce
\begin{equation*}
  \mu_{\rho} = \mu_{Q}
\end{equation*}

\end{document}